\documentclass{IEEEtran}
\usepackage{amsmath}
\usepackage{amssymb}
\usepackage{graphicx}
\usepackage{tabularx}
\usepackage{bm}
\usepackage{cite}
\graphicspath{{./Pictures/}} \DeclareGraphicsExtensions{.pdf,.jpg,.png,.jpeg}

\begin{document}
	
\title{A Markovian influence graph formed from utility line outage data to mitigate large cascades}

\author{Kai Zhou,~\IEEEmembership{Student Member~IEEE}, Ian Dobson,~\IEEEmembership{Fellow~IEEE}, Zhaoyu Wang,~\IEEEmembership{Member~IEEE}\\ Alexander Roitershtein, Arka P. Ghosh\thanks{
Preprint January 2020; to appear in IEEE Transactions on Power Systems.
Manuscript received August 13, 2019; accepted January 26, 2020. 

\copyright 2020 Kai Zhou and Ian Dobson et al. \newline Creative Commons Attribution License CC BY 4.0.

KZ, ID, ZW are with Electrical and Computer Engineering dept.,
Iowa State University, Ames IA USA;
AR is with Statistics dept., Texas A\&M University, College Station TX USA; APG is with Statistics dept., Iowa State University; emails kzhou@iastate.edu, dobson@iastate.edu, wzy@iastate.edu, alexander@stat.tamu.edu, apghosh@iastate.edu.
We gratefully thank BPA for making the outage data public. The analysis and any conclusions are strictly the author's and not BPA's.
We gratefully acknowledge support in part from NSF grants 1609080 and 1735354.

Digital Object Identifier 10.1109/TPWRS.2020.2970406}}
\maketitle

\begin{abstract}
We use observed  transmission line outage data to make a Markovian influence graph that describes the probabilities of transitions between generations of cascading
line outages. Each generation of a cascade consists of a single line outage or multiple line outages.
The new influence graph defines a Markov chain and generalizes previous influence graphs by including multiple line outages as Markov chain states.
The generalized influence graph can reproduce the distribution of cascade size in the utility data. 
In particular, it can estimate  the probabilities of small, medium and large cascades.
The  influence graph  has the key advantage of allowing the effect of mitigations to be analyzed and readily tested, which is not available from the observed data.
We exploit the asymptotic properties of the Markov chain to find the lines most involved in large cascades and show how 
upgrades to these critical lines can reduce the probability of large cascades.
\end{abstract}

\begin{IEEEkeywords}
	cascading failures, power system reliability, mitigation, Markov, influence graph.
\end{IEEEkeywords}

\section{Introduction}
Cascading outages in power transmission systems can cause widespread blackouts.  %\cite{Council2004economic,Arizona,European}
These large blackouts are infrequent, but are high-impact events that occur often enough to pose a substantial risk to society \cite{hinesEP09,carrerasPS16}. 
The power industry has always analyzed specific blackouts and taken steps to mitigate cascading. However, and especially for the largest blackouts of highest risk, the challenges of evaluating and mitigating cascading risk in a quantitative way remain.

There are two main approaches to evaluating cascading risk: simulation and analyzing historical utility data.
Cascading simulations can predict some likely and plausible cascading sequences \cite{baldickPES08,papicPES11}.   However, only a subset of cascading mechanisms can be approximated, and simulations are only starting to be benchmarked and validated for estimating blackout risk \cite{bialekPS16,ciapessoniPMAPS18}.
Historical outage data  can be used to estimate blackout risk \cite{carrerasPS16} and detailed outage data can be used to identify critical lines \cite{papicPMAPS16}.
However it is clear that proposed mitigation cannot be tested and evaluated  with historical data.
In this paper, we process historical line outage data to form a Markovian influence graph that statistically describes the interactions between the observed outages.
The Markovian influence graph can quantify the probability of different sizes of cascades, identify critical lines and interactions, and assess the impact of mitigation on the probability of different sizes of cascades.

\subsection{Literature review}

\looseness=-1
We review the previous literature on influence graphs for power grid cascading outages  and related topics.
There is increasing interest in graphs to represent cascading outages, in which the graph describes the interaction between outaged components and is not the power grid topology. These graphs of interactions have differences in how they are formed and have different names, such as the influence graph, the interaction graph, the correlation network, and the cascading faults graph. The idea of a graph of interactions can be traced back to \cite{AsavathirathamCSM01} which has a stochastic process at each graph node that interacts with different strengths along the graph edges joining to that node to the other nodes. Rahnamay-Naeini \cite{Rahnamay-NaeiniCNC16} generalizes the model of interacting and cascading nodes in \cite{AsavathirathamCSM01} to include interactions within and between two interdependent networks.
This type of interacting particle system model has some nice properties allowing analysis, but remains a somewhat abstract model for power system cascading because it is not known how to estimate the model parameters from data.

Influence graphs in their present form were introduced by Hines and Dobson \cite{hinesHICSS13}, and further developed by Qi, Hines, and Dobson \cite{qiPS15,hinesPS17}.
These influence graphs describe the statistics of cascading data with networks whose nodes 
represent outages of single transmission lines and whose directed edges represent 
probabilistic interactions between successive line outages. 
The more probable edges correspond to the interactions between line outages  that appear more frequently in the data.
Cascades in the influence graph start with initial line outages at the nodes and spread probabilistically along the directed graph edges.
Once the influence graph is formed from the simulated cascading data, it can be used to identify critical components and test mitigation of blackouts by upgrading the most critical components 
\cite{qiPS15,hinesPS17,zhouPMAPS18}.

As well as outages of single lines, cascading data typically includes  multiple line outages that occur nearly simultaneously. 
When the states are single line outages,
these multiple simultaneous outages cause problems  in obtaining well-defined Markov chain transitions between states. 
For example, if the outage of two lines causes an outage in the next generation, it is hard to tell which line caused the subsequent outage or whether the two lines  caused the subsequent outage together.  To address this, \cite{hinesPS17} 
assigns an equal share to the two lines. The resulting influence graph is then  approximated to enable  analysis. Qi \cite{qiPS15} assumes that the subsequent outage is caused by the most frequent line outage.
%, since there is no information about the causal relationship among the outages.  
Improving on this assumption, Qi \cite{qiPS18} considers the causal relationships among successive outages as hidden variables and uses an expectation maximization algorithm to estimate the interactions underlying the multiple outage data.
In this paper, we solve this problem in a novel way by defining an additional state for each 
multiple line outage.  Thus our new influence graph generalizes the interaction between single lines to multiple line outages, so we do not need to make assumptions or approximations when calculating the interactions between two single lines.  This enables a Markov chain to be cleanly and clearly defined.

Considering the different types of  graphs of interactions more generally, there are three methods of quantifying interactions between components which are the edges in the graph of interactions. 
First, as explained in the preceding paragraph, in \cite{hinesHICSS13,hinesPS17,qiPS15}, the edge corresponds to the conditional probability of a single line outage given that the previous line has outaged. 
Second, in \cite{ZhangEPECS13,Merrillpreprint16,MaSG19}, the edge weight is calculated based on the line flow changes due to a single line outage applied to the base case using a DC load flow (In contrast to \cite{hinesHICSS13,hinesPS17,qiPS15} and this paper, this implies that the edge weights do not change during the cascade.). In Merrill \cite{Merrillpreprint16}, the edge weight is obtained from the line outage distribution factors. 
In Zhang \cite{ZhangEPECS13} and Ma \cite{MaSG19}, the directed edge weights are obtained from both the line flow changes and the remaining margin in the line the power is transferred to.
Then Zhang \cite{ZhangEPECS13} combines the directed edges to give undirected edges.
On the other hand, Ma \cite{MaSG19} retains the directed edges and also represents hidden failures by additional nodes.
Third, in Yang \cite{YangPRL17}, the edge corresponds to the correlation between any two lines.  In \cite{CarrerasHICSS12}, Carreras constructs a synchronization matrix from simulation data from the OPA model to identify the lines with higher overloading probabilities.
Other papers \cite{zhouPMAPS18,qiPS18,NakarmiNSE19,JuJESTCS2017,ChenAccess19} form their graph of interactions similarly to  the above methods. 
In this paper, we base the influence graph edges on conditional probabilities. However, the edges are different than the edges in \cite{hinesHICSS13,hinesPS17,qiPS15} as they directly correspond to  transition probabilities in a rigorously defined Markov chain. 

Influence graphs describing the interactions between successive cascading outages were developed using simulated data (Zhou \cite{zhouPMAPS18} is the exception, but \cite{zhouPMAPS18}  differs from this paper because it applies the methods of \cite{hinesPS17} to utility data). But even for simulated cascade data, there remain challenges in extracting good statistics for the influence graph from limited data.
Hines, Dobson and Qi \cite{hinesHICSS13,hinesPS17,qiPS15} estimate the conditional probabilities of transitions with empirical probabilities. In this paper, we mitigate the limited historical cascading data by using a Bayesian method and carefully combining the sparser data of the later stages of cascading in a sophisticated way.

Various measures are proposed for the identification of critical components based on the influence graph.  \cite{hinesPS17,qiPS15,WeiPS18,MaSG19} form their specific measures based on their own influence/interaction graph. Ma \cite{MaSG19} uses a modified page-rank algorithm to find critical lines. Nakarmi \cite{NakarmiNSE19} forms the influence graph using methods of both \cite{hinesPS17} and \cite{YangPRL17},  and proposes a community-based measure to identify critical components. \cite{NakarmiNSE19} compares its measure with other centrality measures based on network theory, and concludes that its method  performs better than other methods in most cases. In this paper, our influence graph is a rigorous Markov chain, and the identification of critical lines is based on the asymptotic quasi-stationary distribution. The quasi-stationary distribution has a clear interpretation of specifying the probabilities that each of the  lines is involved in large cascades.    

The graph of interactions also provides useful information about mitigation actions in power system operation. Ju \cite{JuJESTCS2017} extends the interaction graph to a multi-layer graph, in which the three layers reflect the number of line outages, load shed, and electrical distance of the cascade spread, respectively. This multi-layer graph is suggested to mitigate cascades in system operation by providing the critical lines at different states of cascades. Chen \cite{ChenAccess19} proposes a dynamic interaction graph to better support online mitigation actions than a static interaction graph. During the propagation of a specific cascade, this dynamic interaction graph removes the interactions involving already outaged lines, and optimal power flow controls the power flow on the critical lines indicated by the dynamic interaction graph. The dynamic interaction graph model reduces the risk of large cascades more than the static interaction graph. 

As expected, the graph of interactions and any conclusions drawn depend on the outage data from which the graph is formed. 
If the outage data is simulated, the selection of initial system states matters.
For example, Nakarami  \cite{NakarmiNSE19} shows that different system states lead to different influence graphs. 
In this paper, we form our influence graph from fourteen years of public outage data of a specific area, so that our influence graph reflects the initial faults and states encountered over that period of time in that power system area. The textbook  \cite{Sunbook} includes material on both influence and interaction graphs.

Another related line of research is fault chains.
A fault chain as described in 
\cite{WangPS11} is one cascading sequence of line outages.
Each initial line fault gives a fault chain of lines most stressed at each step until outage or instability.
Usually only the most stressed or most likely next line outage is selected to form fault chains.
By taking each line in the system as the initial outage of each fault chain, Wei \cite{WeiPS18} obtains a set of fault chains using a branch loading index to select the most stressed next line to outage. Each fault chain is expressed as a subgraph whose nodes are transmission lines, and directed edges are branch loading assessment indexes, and the union of the subgraphs forms a cascading faults graph. The edge weights depend on the sum of the branch loading indices, each scaled by the length of the fault they are in.
Then critical lines are identified according to the in- or out-degree of the cascading faults graph. 
Luo \cite{LuoPESGM17} also forms a cascading faults graph with weights depending on load loss in the chain,
and then uses hypertext-induced topic search to select critical lines.
The edge weights of \cite{WeiPS18,LuoPESGM17} differ from those in influence graphs because they are not based on conditional probabilities.
Li and Wu \cite{LiHaoWuPESGM18}
combine simulated fault chains into a network and use reinforcement learning 
to explore, evaluate, and find chains most critical to load shed.
In further work, Li and Wu \cite{LiHaoWuEPSR19}  
combine simulated fault chains into a state-failure network from which 
expected load shed can be computed for each state and failure by 
propagating load shed backwards accounting for the transition probabilities of the edges.
The transition probabilities are estimated similarly to an influence graph by the relative frequency of that transition at that stage of the data. However, in contrast to the practice in influence graphs, the state transition data for the later stages is not combined together to get better estimates. Moreover, fault chains differ from this paper in only 
considering single line outages one after another. 

There are also approaches to modeling cascading with continuous-time Markov processes.
Wang \cite{WangHICSS12} drives line loadings with  generator and load power fluctuations to 
determine overloads and outages that change the Markov state and hence simulate the cascading.
Rahnamay-Naeini \cite{Rahnamay-NaeiniPS14} constructs, using simulated cascading data and fitted functional forms,
a Markov process with states highly aggregated into 3 quantities,
namely the
number of failed lines, the maximum of the capacities of all of the preceding failed lines,  and a cascade stopping index. The aggregated Markov process can model the time evolution of the cascade and the distribution of cascade size.
In further work,
Rahnamay-Naeini reduces the aggregated model to a discrete time Markov chain and generalizes it to model cyber and power  interdependent network cascading interactions in \cite{Rahnamay-NaeiniSG16} and to model operator actions interacting with  cascading in \cite{wangMahshidPS18}.

For another, independent perspective on the literature, Nakarmi's review paper \cite{mahshidArXiv19} surveys various methods of constructing interaction graphs and the reliability analysis based on interaction graphs.

\subsection{Contributions of paper} 

The new influence graph generalizes and improves previous  work in several ways.
In particular, this paper
\begin{itemize}
\item uses real data observed and routinely collected by utilities rather than simulated data.
\item obtains a clearly defined influence graph that solves
 the problem of multiple simultaneous outages by using additional states with multiple outages. This generalized influence graph 
 rigorously defines a Markov chain.
 \item mitigates the problems of limited cascading data with several new methods; in particular, it combines Bayesian methods of estimation with elaborate methods of distinguishing and combining different events. This better estimates the transition matrices of the influence graph while matching the increasing cascade propagation and retaining possibilities of analysis.
   \item computes the probabilities of small, medium and large cascades, and these match the historical data statistics.
  \item makes innovative use of the bootstrap to estimate the variance of the probabilities of small, medium and large cascades. This allows checking that the estimated probabilities of small, medium and large cascades are accurate enough to be useful.
    \item calculates critical lines most involved in large cascades directly from the Markov chain as the quasi-stationary distribution of the Markov chain.   
\end{itemize}
All of these advances clearly distinguish this paper from the previous work reviewed above.

\section{Forming the Markovian influence graph from historical outage data}
\label{sec:forming}

We use detailed historical line outage data consisting of records of individual automatic transmission line outages that specify the lines outaged and the outage times to the nearest minute. We emphasize that this data is routinely recorded by utilities worldwide, for example in the North American Transmission Availability Data System.

The first step in building an influence graph is to take many cascading sequences of transmission line outages and divide each cascade\footnote{The grouping of line outages into cascades uses the simple method of \cite{dobsonPS12}:
 The grouping  is done by looking at the gaps in start time between successive line outages.	If successive outages have a gap of one hour or more, then the outage after the gap starts a new cascade. More elaborate methods of grouping real line outages into cascades could be developed and applied.}
into generations of outages 
as detailed in \cite{dobsonPS12}.
Each cascade starts with initial line outages in generation 0, and continues with subsequent generations of line outages 1,2,3,... until the cascade stops. Each generation of line outages is a set of line outages that occur together on a fast time scale of less than one minute. 
Often there is only one line outage in a generation, but protection actions can act quickly to cause several line outages in the same generation.
(Sometimes in a cascading sequence an outaged line recloses and outages in a subsequent generation.
In contrast to \cite{dobsonPS12,zhouPMAPS18}, here we neglect the repeats of these outages.)

\begin{figure}[ht]
	\centering
	\includegraphics[width=0.95\columnwidth]{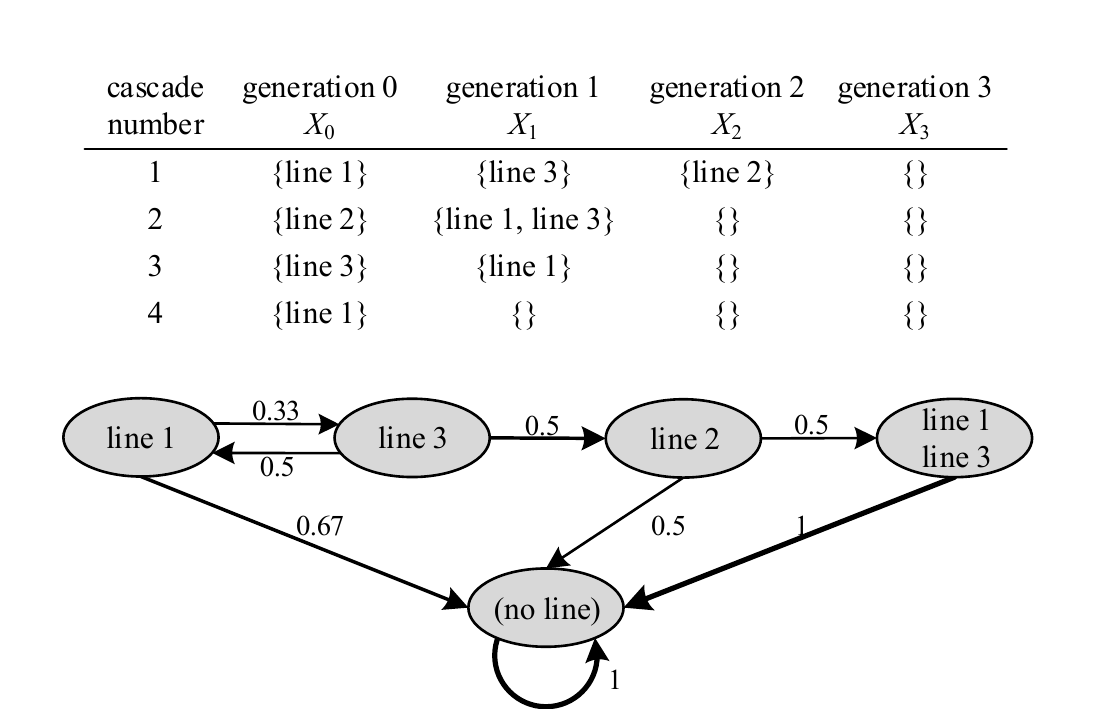}
	\caption{Simple  example forming influence graph from artificial data (real utility data is shown in Fig.~\ref{fig:influencegraph}).}
	\label{fig:egmig}
\end{figure}

\begin{figure}[!t]
	\centering
	\vspace{-0mm}
	\rotatebox[origin=c]{-90}{\includegraphics[width=1.2\columnwidth]{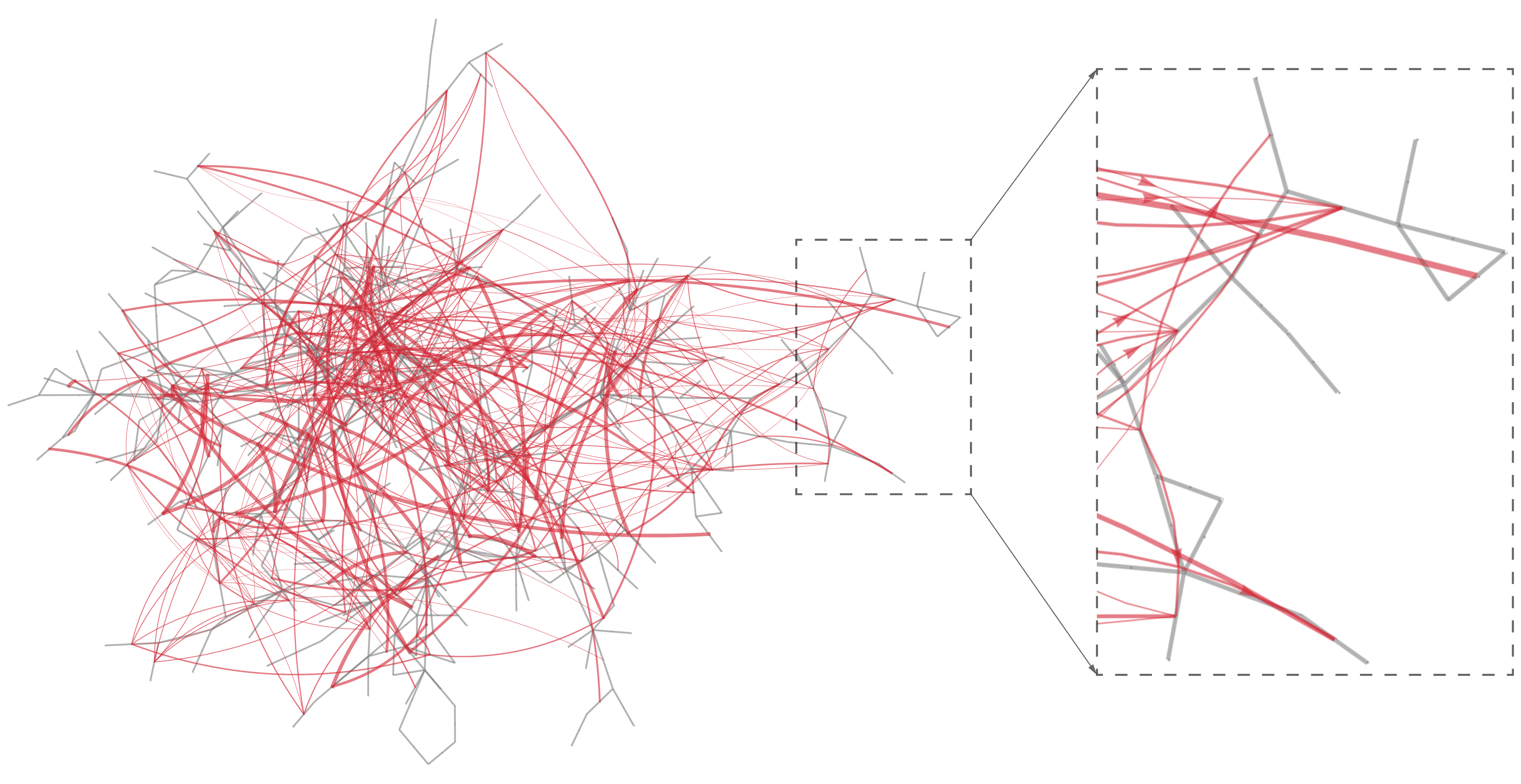}}
	\vspace{-1mm}
	\caption{The gray network is the system network and the red network is the influence graph showing the main influences between lines. The  red edge thickness indicates the strength of the influence. 
	}
	\label{fig:influencegraph}	
\end{figure}

The influence graph represents cascading as a Markov chain   $X_0, X_1,... $, in which $X_k$ is the set of line outages in generation $k$ of the cascade.  
We first illustrate the formation of the influence graph from artificial cascading data with the simple example of four observed cascades involving three lines shown in Fig.~\ref{fig:egmig}.
The first cascade has line 1 outaged in generation 0, line 3 outaged in generation 1, line 2 outaged in generation 2, and then the cascade stops with no lines (indicated by the empty set \{\}) outaged in generation 3.
All cascades eventually stop by transitioning to and remaining in the state \{\} for all future generations.
%The second cascade has lines 1 and 3 outaging together in generation 1.
The five states observed in the data are  \{\}, \{line 1\},  \{line 2\}, \{line 3\}, and \{line 1, line 3\}, where this last state is lines 1 and 3 outaging together in the same generation, as in generation 1 of cascade 2. Introducing the state  \{line 1, line 3\} with two line outages avoids the problems in previous work in accounting for transitions to and from the simultaneous outages of  line 1 and  line 3.

We can estimate the probabilities of transitioning from state $i$ to state $j$ in the next generation by counting the number of those transitions in all the cascades and dividing by the number of occurrences of state $i$.
For example, the probability of transitioning from state \{line 1\} to state \{line 3\} is $1/3$ and the probability of transitioning from state \{line 2\} to state \{line 1, line 3\} is $1/2$. 
The probability of transitioning from state \{line 1\} to \{\}; that is, stopping
after the single outage of line 1,
is $2/3$.
The probabilities of the edges out of each state sum to 1.
By working out all the transition probabilities, we can make the network graph of the Markov chain as shown in Fig.~\ref{fig:egmig}.
The transitions between states with higher probability are shown with thicker lines.
In this generalized influence graph, the nodes are sets of line outages and the edges indicate transitions or interactions between sets of line outages in successive generations of cascading. The influence graph is different than the physical grid network and cascades are generated in the influence graph by moving along successive edges, selecting them according to their transition probabilities.

In the general case, there are many states $s_0$, $s_1$, ... , and we describe the transitions between them.
Let $\bm P_k$ be the Markov chain transition matrix for generation $k$. The $\bm P_k$ matrix entry $P_k[i,j]$ is the conditional probability that the set of outaged lines is $s_j$ in generation $k+1$, given that the set of outaged lines is $s_i$ in generation $k$; that is,
\begin{align}
	P_k[i,j]={\rm P}[X_{k+1}= s_j~|~X_{k}=s_i].
	\label{Pk}
\end{align} 

The key task of forming the Markov chain is to estimate the transition probabilities in the matrix $\bm P_k$ from the cascading data. 
If one supposed that $\bm P_k$ does not depend on $k$, a straightforward way to do this would first construct a counting matrix $\bm N$ whose entry $N[i,j]$ is the number of transitions from $s_i$ to $s_j$ among all generations in all the cascades.  Then $\bm P_k$ would be estimated as  
\begin{align}
P_k[i,j]=\frac{N[i,j]}{\sum_{j}N[i,j]}.
\label{PkN}
\end{align}
However, we find that $\bm P_k$ must depend on $k$ in order to reproduce the increasing propagation of outages observed in the data
\cite{dobsonPS12}. On the other hand, there is not enough data to accurately estimate $\bm P_k$ individually for each $k>0$.
Our solution to this problem involves both grouping together data for higher generations and having $\bm P_k$ varying with $k$, as well as 
using empirical Bayesian methods to improve the required estimates of cascade stopping probabilities.
The detailed explanation of this solution is postponed to section~\ref{estimateP}, and until  section~\ref{estimateP} we assume that $\bm P_k$ has already been estimated 
for each generation $k$ from the utility data.
Forming the Markov chain transition matrix from the data in this way makes the Markovian assumption that the statistics of the 
lines outaged in a generation only depend on the lines outaged in the previous generation. 
This is a pragmatic assumption that yields a tractable data-driven probabilistic model of cascading.

One way to visualize the influence graph interactions between line outages in $\bm P_k$ is to restrict attention to the interactions between single line states, and show these as the red network in Fig.~\ref{fig:influencegraph}. The gray network is the actual grid topology, and the gray transmission lines are 
joined by a red line of the thickness proportional to the probability of being in successive generations, if that probability is sufficiently large.
The interactions in Fig.~\ref{fig:influencegraph} reflect a very wide range of mechanisms. 
The longer-range mechanisms include redistributions of power due to line and generator outages, remedial action schemes, and bad weather across the grid.

Let the row vector $\bm \pi_k$ be the probability distribution of states in generation $k$. The  $\bm \pi_k$ entry $\pi_k[i]$  is the probability that the set of outaged lines is $s_i$ in generation $k$; that is,
\begin{align}
\pi_k[i]= {\rm P}[X_k = s_i].
\end{align} 
Then the propagation of sets of line outages from generation $k$ to generation $k+1$ is given by
\begin{align}
\bm \pi_{k+1} = \bm \pi_{k} \bm P_k
\label{markov}
\end{align} 
and, using (\ref{markov}),  the distribution of states in generation $k$ depends on the initial distribution of states  $\bm \pi_0$ according to
\begin{align}
\bm \pi_k  = \bm \pi_{0} \bm P_{0} \bm P_{1} ... \bm P_{k-2} \bm P_{k-1}.
\label{equ:pik}
\end{align} 

\section{Illustrative historical outage data}
\label{sec:illustrative}

While our method applies generally to the detailed outage data routinely collected by utilities, we illustrate our method with a specific publicly available data set, which is the automatic transmission line outages recorded by a large North American utility over 14 years starting in 1999 \cite{bpadata}. 
We group the 9,741 line outages into 6,687 cascades \cite{dobsonPS12}. Most of the cascades (87\%) have one generation because initial outages often do not propagate further.
There are 614 lines and the observed cascades have 1094 subsets of these lines that form the 1094 states 
$s_0$, $s_1$, ... , $s_{1093}$.  Among these 1094 states, 50\% have multi-line outages. And among these multi-line outage states, about 20\% are comprised of lines sharing no common buses.
While in theory there are  $2^{614}$  subsets of  614 lines, giving an impractically large number of states, we find 
in practice with our data that the number of states is less than twice the number of lines. Note that our statistical modeling approximates the power grid as unchanging over the time span of the data \cite{dobsonPS16}. In practice a utility would have the records of changes to partially mitigate the effects of this approximation.

\section{Computing the distribution of cascade sizes and its confidence interval}

We compute the distribution of cascade sizes from the Markov chain and check that it reproduces the empirical distribution of cascade sizes,
and estimate its confidence interval with a bootstrap.

We can measure the cascade size by its number of generations. 
Define the survival function of the number of generations in a cascade as
\begin{align}
S(k) = {\rm P}[\mbox{number of cascade generations $>k$}]
\label{equ:sf} 
\end{align}
$S(k) = 1 - \pi_{k}[0]$, where
$\pi_{k}[0]$ is the probability that a cascade is in state $s_0=\{\}$ in generation $k$ and also the probability that the cascade stops at or before generation $k$. Hence 
\begin{align}
S(k) &= 1 - \pi_{k}[0] =\bm \pi_{k} ({\bf1} -\bm e_0)  \notag \\
&=\bm \pi_{0} \bm P_{0} \bm P_{1} ... \bm P_{k-2} \bm P_{k-1} ({\bf1} -\bm e_0),
\label{equ:sff} 
\end{align}
where ${\bf1}$ is the column vector $(1,1,1,...,1)'$, and $\bm e_0$ is the column vector $(1,0,0,0,...,0)'$.
The initial state distribution $\bm\pi_{0}$ can be estimated directly from the cascading data.

\begin{figure}[ht]
	\centering
	\includegraphics[width=0.9\columnwidth]{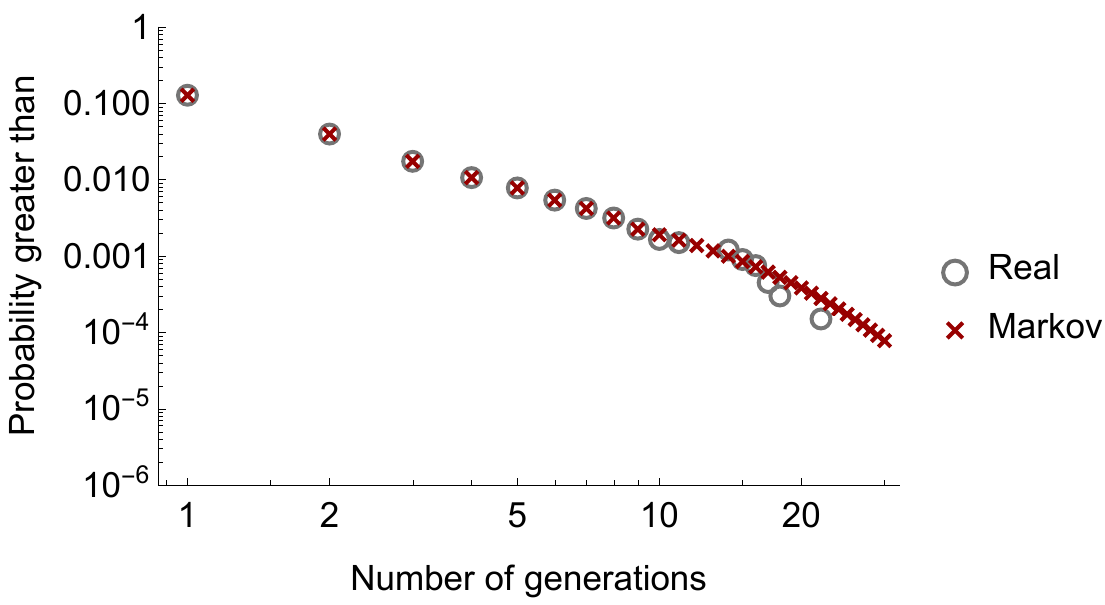}
	\caption{Survival functions of the number of generations from real data and from the Markov chain.}
	\label{fig:sizedistSF}
\end{figure}

Then we can confirm that the influence graph reproduces the statistics of cascade size in the cascading data by 
comparing the survival function  $S(k)$ computed from (\ref{equ:sff}) with the 
empirical survival function computed directly from the cascading data as shown in Fig.~\ref{fig:sizedistSF}.
The Markov chain reproduces the statistics of cascade size closely, with a Pearson $\chi^2$ goodness-of-fit test $p$-value of $0.99$.

We use bootstrap resampling \cite{davisonbootstrap97} to estimate the variance of our estimates of probabilities of cascade sizes.
A bootstrap sample resamples the observed cascades with replacement, reconstructs the Markov chain, and recomputes the probabilities of cascade sizes. Note that each bootstrap resampling amounts to a different selection of the cascades observed in the data.
The variance of the probabilities of cascade sizes is then obtained as the empirical variance of the bootstrap samples. We use 500 bootstrap samples to ensure a sufficiently accurate estimate of the variance of the probabilities.

\begin{figure}[ht]
	\centering
	\includegraphics[width=0.9\columnwidth]{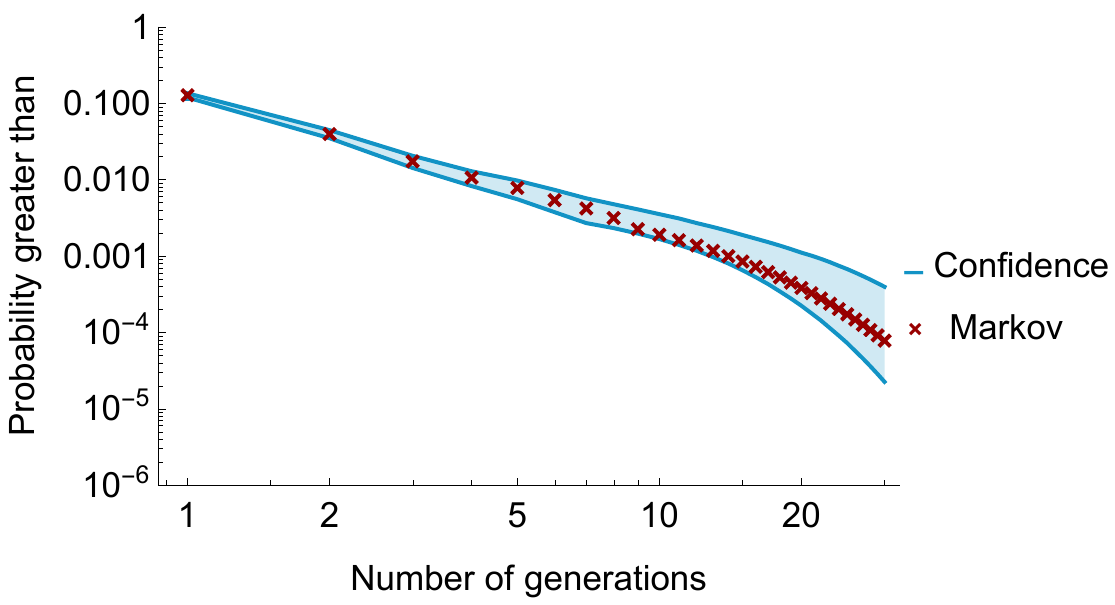}
	\caption{Survival function of cascade sizes. Red crosses are from Markov chain, and blue lines indicate the 95\% confidence interval estimated by bootstrap.}
	\label{fig:bootstrapsfs}
\end{figure}

The risk of a given size of blackout is estimated as 
risk = (estimated probability $\hat p$ of that size of blackout) $\times$ (cost of that size of blackout).
Knowing the multiplicative uncertainty in $\hat p$ is useful.
 For example, if we know $\hat p$ to within a factor of  2, then this contributes a factor of
2 to the uncertainty of the risk.
Therefore, it is appropriate to use a multiplicative form of confidence interval for $\hat p$ 
specified by a parameter~$\kappa$.  A 95\% multiplicative confidence interval for an estimated probability $\hat p$ means that the probability $p$ satisfies ${\rm P}[\hat p/ \kappa \leq p \leq \hat p\, \kappa] = 0.95$.
The confidence interval for the estimated survival function is shown in Fig.~\ref{fig:bootstrapsfs}. Since larger cascades are rarer than small cascades, the variation increases as the number of generations increases.  

To apply and communicate the probability distribution of cascade size, it is convenient to combine sizes together to  get the 
probabilities of small, medium, and large cascades, where a small cascade has 1 or 2 generations,
a medium cascade has 3 to 9 generations, and a large cascade has 10 or more generations.
(The respective probabilities are calculated as $1-S(2)$, $S(2)-S(9)$, and $S(9)$).
The 95\% confidence intervals of the estimated probabilities of small, medium, and large cascades are shown in Table \ref{tbl:bootstrapnoloop}.
The probability of large cascades is estimated within a factor of 1.5, which is adequate for the purposes of 
estimating large cascade risk, since the cost of large cascades is so poorly known: estimates of the direct costs of cascading blackouts 
 vary by more than a factor of 2. 

\begin{table}[ht]
	\centering
	\caption{95\% Confidence intervals using bootstrap}
	\label{tbl:bootstrapnoloop}
	\begin{tabular}{ccc}
		cascade size & probability  &  $\kappa$ \\
		\hline
		small (1 or 2 generations)  & 0.9606  &  1.005\\
		medium (3 to 9 generations)  & 0.0372  &  1.132\\
		large (10 or more generations)  & 0.0022  &   1.440\\
		\hline		
	\end{tabular}
\end{table}

\looseness=-1
We now discuss tracking cascades by their number of generations.
The number of generations is the same concept as the number of tiers in commercial cascading software \cite{VaimanPS12}.
Basic to cascading analysis is the grouping of line outages into successive generations within each cascade. This grouping is usually done by outage timing as in this paper, or by simulation loops naturally producing generations of outages.
This paper is structured in terms of these generations, so that propagation is determined by the probability of a next generation (i.e. the cascade not stopping at the current generation), and cascade size is measured by number of cascade generations. In contrast, some previous papers \cite{dobsonPS12,hinesPS17,zhouPMAPS18,papicPMAPS16} are structured in terms of the line outages in the generations, so that, according to the branching process model \cite{dobsonPS12}, each line outage in each generation propagates independently to form line outages in the next generation. Then the propagation is determined by the number of line outages per line outage in the previous generation,  and it is natural to use the total number of lines outaged as a measure of cascade size.
While it is not yet clear which approach is better, there may be some advantages to tracking cascades by generations rather than line outages.  Generations are simpler and more general than line outages, and can more easily encompass other outages significant in cascading such as transformer outages. Also, it may be that the statistics of the number of generations is more simply described, as in the Zipf distribution observed in utility data in \cite{DobsonArXiv18}.

\section{Critical lines and cascade mitigation}

\subsection{The transmission lines involved in large cascades}
\label{sec:involved}

The lines eventually most involved in large cascades can be calculated from the asymptotic properties of the Markov chain.
While all cascades eventually stop, we can consider at each generation those propagating cascades that are not stopped at that generation.
The probability distribution  of states involved in these propagating cascades converges to a probability distribution $\bm d_\infty$, which is called the quasi-stationary distribution.
$\bm d_\infty$ can be computed directly from the transition matrices
(as explained in Appendix \ref{app3}, $\bm d_\infty$ is the left eigenvector corresponding to the dominant eigenvalue of the 
transition submatrix $\bar{\bm Q}_{1+}$). That is, except for a transient that dies out after some initial generations, the participation of states in the cascading that continues past these initial generations is well approximated by $\bm d_\infty$.
Thus the high probability states corresponding to the highest probability entries in $\bm d_\infty$ are 
the critical states most involved in the latter portion of large cascades. Since $\bm d_\infty$ does not depend 
on the initial outages, the Markov chain is supplying information about the eventual cascading for all initial outages.

\begin{figure}[ht]
	\centering
	\includegraphics[width=0.8\columnwidth]{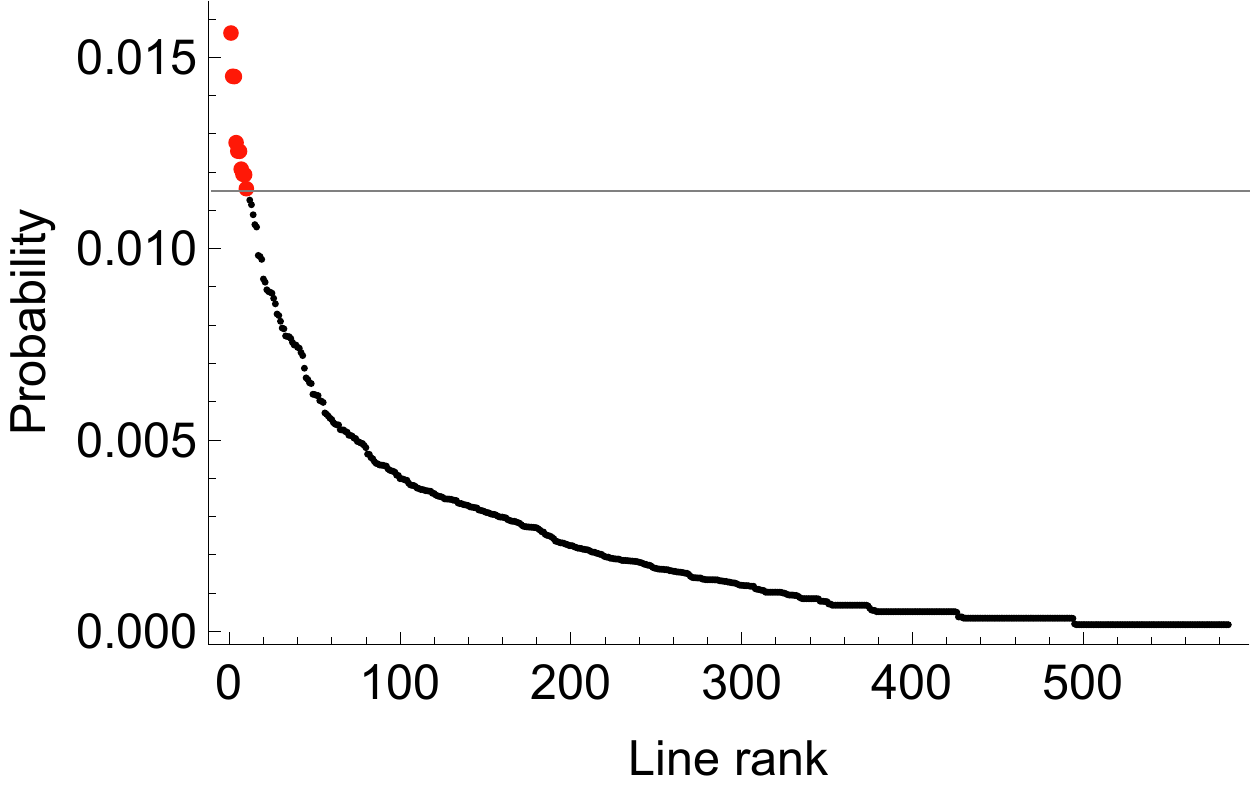}
	\caption{Quasi-stationary distribution of transmission lines eventually involved in propagating cascades. Red dots are ten critical lines.}
	\label{fig:criticalLines}
\end{figure}

We now find the critical lines corresponding to these critical states by projecting the states onto the lines in those states.
Let $\bm \ell_k$ be the row vector whose entry $\ell_k[j]$ is the probability that line $j$ outages in generation $k$. 
Then
\begin{align}
\ell_k[j]= \sum_{i: j \in s_i} \pi_k[i]\quad\mbox{or}\quad \bm \ell_k= \bm \pi_k  \bm R,
\end{align}
where the matrix $ \bm R$ projects states to lines according to 
\begin{align}
R[i,j]=\left\{
\begin{array}{lr}
1;& \mbox{line}~ j \in s_i\\
0;&\mbox{line}~ j \notin s_i
\end{array}
\right.
\end{align}
Then the probability distribution  of lines eventually involved in the propagating cascades that are not stopped is $\bm c_\infty=\bm d_\infty \bm R$
and the critical lines most involved in the latter portion of large cascades correspond to the highest probability entries in $\bm c_\infty$.
Fig.~\ref{fig:criticalLines} shows the probabilities in $\bm c_\infty$ in order of decreasing probability. We identify the top ten lines as critical and as candidates for upgrading to decrease the probability of large cascades.

\subsection{Modeling and testing mitigation in the Markov chain}

A transmission line is less likely to fail due to other line outages after the line is upgraded, its protection is improved, or its operating limit is reduced.
These mitigations have the effect of decreasing the probability of transition to states containing the upgraded line, and are an adjustment 
of the columns of the transition matrix corresponding to these states.
The mitigation is represented in the Markov chain by reducing the probability of transition to the state $s$ containing  the upgraded line by $({r}/{|s|})\%$, where $|s|$ is the number of lines in the state.
The reduction is $r\%$ if the state contains only the upgraded line, and the reduction is less if the state contains multiple lines.

\begin{figure}[ht]
	\centering
	\includegraphics[width=0.9\columnwidth]{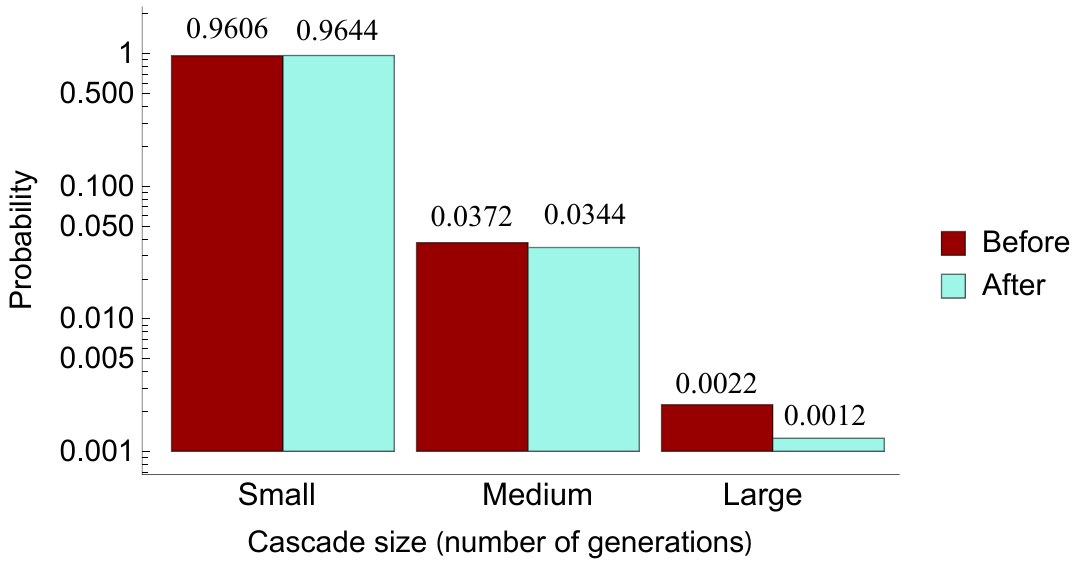}
	\caption{Cascade size distribution before (red) and after (light green) mitigating lines critical in propagating large cascades.}
	\label{fig:mitigationProBar}
\end{figure}

We demonstrate using the Markov chain to quantify the impact of mitigation by 
upgrading the ten lines critical for large cascades identified in section~\ref{sec:involved} with $r=80\%$.
%We test mitigation by upgrading ten critical lines. It will effectively decrease the probability of large cascades. %It is because large cascades have long generations so they are exposed more time to mitigation effect. 
The effect of this mitigation on cascade probabilities is shown in Fig.~\ref{fig:mitigationProBar}. It shows that upgrading the critical lines
reduces the probability of large cascades by 45\%, while the probability of medium cascades is slightly decreased and the probability of small cascades is slightly increased.% because some large cascades become small cascades after mitigation.  

To show the effectiveness of the method of identifying critical lines, we compare the mitigation effect of upgrading critical lines and upgrading ten random lines. 
Randomly upgrading ten lines only decreases the probability of large cascades by 11\% on average.

\looseness=-1
So far we have only considered upgrading the lines critical for propagating large cascades.
Now, in order to discuss this mitigation of large cascades in a larger context, we briefly consider and contrast
a different mitigation tactic of upgrading lines that are critical for initial outages.
Since initial outages are caused by external causes such as storm, lightning, or misoperation, they often have different mechanisms and different mitigations than for propagating outages. A straightforward method to identify lines critical for initial outages selects the ten lines  in the data with the highest frequencies of initial outage \cite{zhouPMAPS18}. 
Upgrading these ten lines will reduce their initial outage frequencies and hence reduce the overall cascade frequency. 
In the Markov chain, this upgrading is represented by reducing in the first generation the frequency of states $s$ that contain the critical lines for initial outages by $r/|s|$\%, where $r=80\%$.
The main effect is that by reducing the initial outage frequencies of the critical lines by 80\%, we reduce the frequency of all cascades by 19\%.
In addition, this mitigation will change the probabilities of states $\bm \pi_0$ after renormalizing the frequencies of states.
It turns out for our case that there is no overlap between critical lines for initial outages and for propagation.

Changing the initial state distribution $\bm \pi_0$ has  no effect on the 
distribution of cascade sizes in the long-term. However, it directly reduces the frequency of all cascades.
In contrast, mitigating the lines critical for propagating large cascades reduces the probability of large cascades relative to all 
cascades but has no effect on the frequency of all cascades.
(Note that Fig.~\ref{fig:mitigationProBar} shows the distribution of cascade sizes assuming that there is a cascade, but gives no information about the frequency of all cascades.)

In practice, a given mitigation measure can affect both the initial outages and the propagation of outages into large cascades.
The combined mitigation effects can also be represented in the influence graph by changing both the initial state distribution and the transition matrix, but here it is convenient to discuss them separately.

This paper aims to select the lines critical for large cascades and quantify  the impact on cascade probability of  generic upgrades to these lines.
Once the critical lines are selected,  an engineering process of much wider scope is required to determine the possible approaches to upgrade each of the lines, quantify the benefits other than reducing large cascades and balance the costs and feasibilities of the upgrading approaches against the total benefits of upgrading.
One part of this process is that for each line,
the percentage reduction in outage probability for the best approach to line upgrade is estimated and the 
Markov chain is used to quantify the corresponding reduction in large, medium, and small cascade probabilities.
However, cascade mitigation is only one of the many factors to be considered in justifying upgrade.
Evaluating and costing specific upgrading approaches for specific lines requires utility expertise, including details of the line construction and right of way, maintenance history, and operation.

\section{Estimating the transition matrix}
\label{estimateP}

The Markov chain has an absorbing first state $s_0=\{\}$, indicating no lines outaged as the cascade stops and after the cascade stops.
Therefore the transition matrix has the structure 
\begin{align} \bm P_k = 
\left[ {\begin{array}{*{20}{c}}
	1&\vline& 0& \cdots &0\\
	\hline
	{}&\vline& {}&{}&{}\\
	\bm u_k&\vline&{}&\bm Q_k&{}\\
	{}&\vline&{}&{}&{}\\
	\end{array}} \right]
\end{align}   
where $\bm u_k$ is a column vector of stopping probabilities; that is, $u_k[i] = P_k[i,0]$.
 $\bm Q_k$ is a submatrix of transition probabilities between transient states which contains the non-stopping probabilities. 
 The first row of $\bm P_k$ is always $\bm e'_0$, so the transition probabilities to be estimated are  $\bm u_k$ and 
 $\bm Q_k$ for each generation $k$. The rows and columns of $\bm P_k$ are indexed from 0 to $|\mathcal{S}|-1$ and the rows and columns of $\bm Q_k$
 are indexed from 1 to $|\mathcal{S}|-1$, where $|\mathcal{S}|$ is the number of states.
 
As summarized in section \ref{sec:forming} after (\ref{Pk}), we need to both group together multiple generations to get sufficient data and account for variation with generation $k$.
The statistics of the transition from generation 0 to generation 1 are different than the statistics of the transitions between the subsequent generations. For example, stopping probabilities for generation 0 are usually larger than stopping probabilities for subsequent generations \cite{zhouPMAPS18}.
Also, the data for the subsequent generations is sparser.
Therefore, we %use (\ref{PkN}) to 
construct from counts of the number of transitions from generation 0 to generation 1 a probability transition matrix $\bar{\bm P}_0$, and construct
from the total counts of the number of transitions from all the subsequent generations a probability transition matrix $\bar{\bm P}_{1+}$. Specifically, we first use the right-hand side of  (\ref{PkN}) to construct two corresponding empirical transition matrices, and then we update stopping probabilities by the empirical Bayes method and adjust non-stopping probabilities to obtain $\bar{\bm P}_0$ and $\bar{\bm P}_{1+}$. Finally, we adjust $\bar{\bm P}_0$ and $\bar{\bm P}_{1+}$ to match the observed propagation rates to obtain $\bm P_k$ for each generation $k$.

\subsection{Bayesian update of stopping probabilities}
\label{subsec:updating}

The empirical stopping probabilities are improved by an empirical Bayes method  \cite{Guikema2007Formulating,carlinBayesian08} to help mitigate the sparse data for some of these probabilities.    
Since the method is applied to both  $\bar{\bm P}_{0}$ and $\bar{\bm P}_{1+}$, we simplify notation     by writing  $\bar{\bm{P}}$ for
either $\bar{\bm P}_{0}$  or $\bar{\bm P}_{1+}$. 

The matrix of empirical probabilities obtained from the transition counts $N[i,j]$ is
\begin{align}
{\bar P}^{\rm counts}[i,j]=\frac{N[i,j]}{\sum_{j}N[i,j]}
\label{PN}
\end{align}

We construct $\bar{\bm{P}}$ from $\bar{\bm{P}}^{\rm counts}$  in two steps. First,  Bayesian updating is used to better estimate stopping probabilities and form a matrix $\bar{\bm{P}}^{\rm bayes}$. 
Second, the non-stopping probabilities in $\bar{\bm{P}}^{\rm bayes}$ are adjusted to form the matrix $\bar{\bm{P}}$ to account for the fact that some independent outages are grouped into cascading outages when we group outage data into cascades.

We need to estimate  the probability of the cascade stopping at the next generation for each state encountered in the cascade.
For some of the states, the  stopping counts  are low, and cannot give good estimates of the stopping probability.
However, by pooling the data for all the states we can get a good estimate of the mean probability of stopping over all the states.
We use this mean probability to adjust the sparse counts in a conservative way.
In particular, we form a prior that maximizes its entropy subject to the mean of the prior being the mean of the pooled data.
This maximum entropy prior can be interpreted as the prior distribution that makes the least 
possible further assumptions about the data \cite{jaynes86}\cite{fangEntropy12}.

\paragraph{Finding a maximum entropy prior}

Assuming the stopping counts are independent with a common probability, the stopping counts follow a binomial distribution. Its conjugate prior distribution is the beta distribution, whose parameters are estimated using the maximum entropy method. 

Let stopping counts $C_i$ be the observed number of transitions from state $s_i$ to $s_0$ ($i = 1,...,|\mathcal{S}|-1$). Then $C_i = N[i,0]$. Let $n_i = \sum_{j=0}^{|\mathcal{S}|-1} N[i,j]$ be the row sum of the counting matrix $\bm N$.
The stopping counts $C_{i}$ follow a binomial distribution with parameter $U_i $, with probability mass function 
\begin{align}
f_{C_i|U_i}(c_i | u_i) & = \dfrac{n_i !}{c_i ! (n_i-c_i)!}  u_i^{c_i} (1-u_i)^{n_i - c_i} \label{equ:pdfBin}
\end{align}
The conjugate prior distribution for the binomial distribution is the beta distribution. 
Accordingly, we use  the beta distribution with hyperparameters $\beta_1, \beta_2$ for the stopping probability $ U_i $:
\begin{align}
f_{U_i}(u_i) =B(\beta_1,\beta_2) u_i^{\beta_1 -1} (1-u_i)^{\beta_2 - 1}  \label{equ:priorBeta}
\end{align} 
where $B(\beta_1,\beta_2)= \frac{\Gamma(\beta_1+\beta_2)}{\Gamma(\beta_1) \Gamma(\beta_2)} $.
Alternative parameters for the beta distribution are  its precision $m = \beta_1 + \beta_2$ and its mean  $\mu = \frac{\beta_1}{\beta_1 + \beta_2}$. 
The entropy of the beta distribution is 
\begin{align}
&{\rm Ent}(m,\mu)= \ln B(m \mu,m(1-\mu)) - (m \mu-1) \psi (m \mu)  \notag \\
& \qquad - (m (1-\mu)-1) \psi (m (1-\mu))  + (m - 2) \psi (m) \label{equ:maxent} 
\end{align}
where $\psi(x)$ is the digamma function.

We want to estimate hyperparameters $\beta_1$, $\beta_2$ to make the beta distribution have maximum entropy subject to the mean being the average stopping probability of the pooled data
$\hat{u} = (\sum_{i=1}^{|\mathcal{S}|-1} c_{i})/(\sum_{i=1}^{|\mathcal{S}|-1}  n_i  )$.
Then we can obtain hyperparameters  $\beta_1 $,  $\beta_2$ by finding the $m>0$ that maximizes ${\rm Ent} (m,\hat u)$ and 
evaluating  $\beta_1 =  m \hat u $ and  $\beta_2 = m (1-\hat u)$.
The hyperparameters used  for $\bar{\bm P}_{0}^{\rm bayes}$ are $(\beta_1, \beta_2) = (2.18, 0.32)$, and the hyperparameters for $\bar{\bm P}_{1+}^{\rm bayes}$ are $(\beta_1, \beta_2) = (1.10, 0.93)$.

\begin{figure}[t!]
	\centering
	\includegraphics[width=0.9\columnwidth]{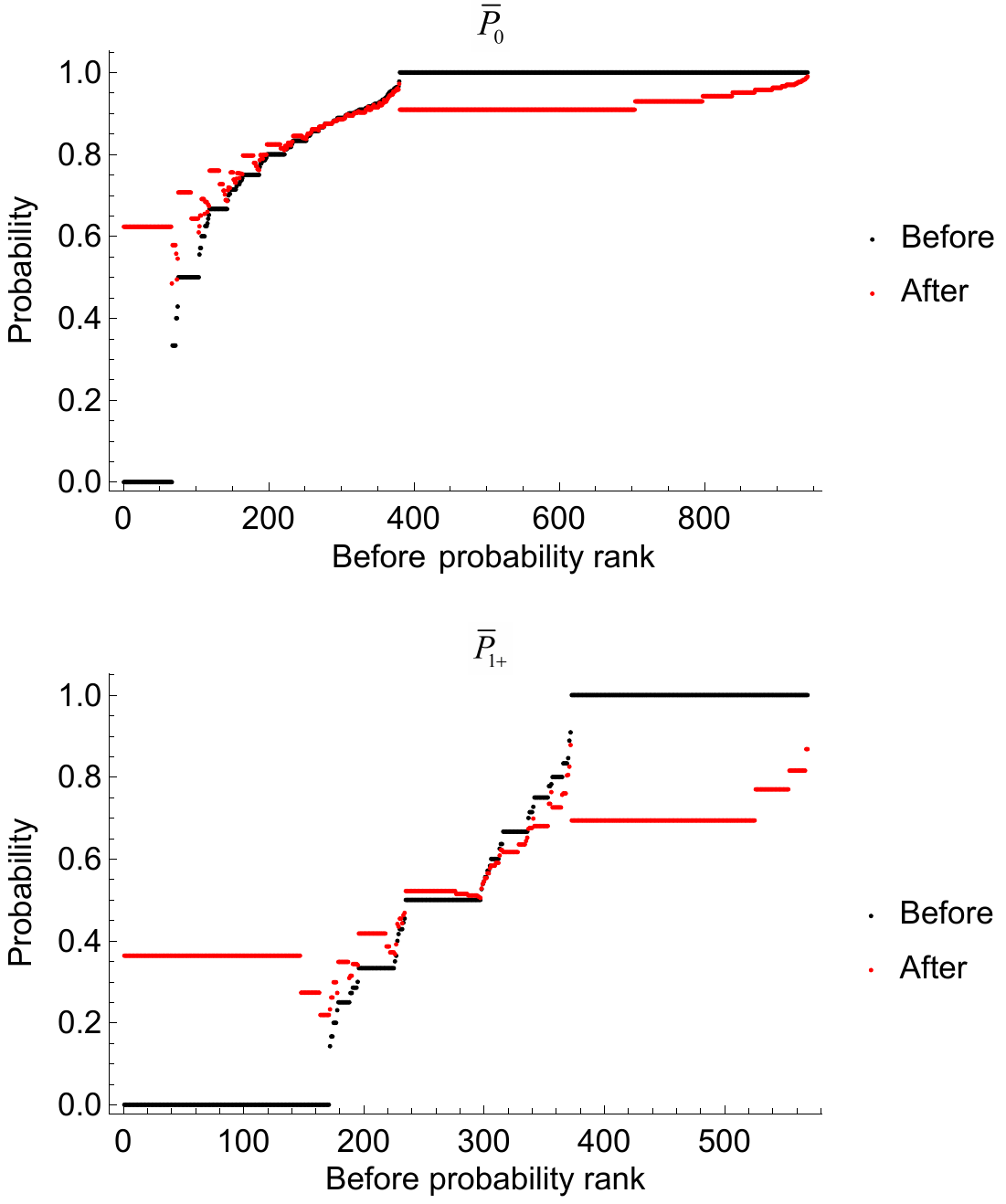}
	\caption{Stopping probabilities before and after Bayesian updating}
	\label{fig:compProbStop}
	%\vspace{-5mm}
\end{figure}

\paragraph{Updating the observed data using the prior}
The posterior distribution of the stopping probability $U_i$ is
a beta distribution with parameters $c_{i} + \beta_1, n_i - c_{i} + \beta_2$. We use the mean of the posterior distribution as a point estimate of the stopping probability:
\begin{align}
\bar{P}^{\rm bayes}[i,0]  = {\rm E}( U_i | C_{i} = c_{i}) = \dfrac{c_{i} + \beta_1}{n_{i} + \beta_1 + \beta_2}  
 \label{equ:expPostBin}
\end{align} 
Fig.~\ref{fig:compProbStop} shows a comparison between the empirical stopping probabilities and the updated  stopping probabilities. Black dots are the empirical probabilities sorted in ascending order (if two probabilities are equal, they are sorted according to the total counts  observed). Red dots are the updated stopping probabilities. As expected, the empirical probabilities with the fewest counts move towards the mean the most 
when updated. As the counts increase, the effect of the prior decreases and the updated probabilities tend to the empirical probabilities. 

Equation (\ref{equ:expPostBin}) forms the first column of $\bar{{\bm P}}^{\rm bayes}$.
Then the nonstopping probabilities in the rest of the columns of the ${\bar{\bm P}}^{\rm counts}$ matrix are scaled so that they sum to one minus the stopping probabilities of  (\ref{equ:expPostBin}) to complete the matrix $\bar{{\bm P}}^{\rm bayes}$:
\begin{align}
\bar{P}^{\rm bayes}[i,j]  =\frac{1-\bar{P}^{\rm bayes}[i,0] }{\sum_{r=1}^{|\mathcal{S}|-1}\bar{P}^{\rm counts}[i,r]} \bar{P}^{\rm counts}[i,j] ,\  j>0
 \label{equ:nonstopadjust}
\end{align} 
This Bayesian updating is applied to form  $\bar{\bm P}_{0}^{\rm bayes}$ for the first transition and  $\bar{\bm P}_{1+}^{\rm bayes}$ for the subsequent transitions.

\subsection{Adjust nonstopping probabilities for independent outages}

The method explained in section  \ref{sec:forming} that groups outages into cascades has an estimated 6\% chance that it groups independent outages into cascading outages \cite{dobsonPS16}. These 6\% of outages occur independently while the cascading of other outages proceeds and do not arise from interactions with other outages. 
The empirical data for the nonstopping probabilities includes these 6\% of outages, and we want to correct this.
Therefore, the non-stopping probabilities are modified by shrinking the probabilities in transition matrix by 6\%, and sharing this equally among all the states. That is,
\begin{align}
	\bar{P}[i,j] = 0.94 \bar{P}^{\rm bayes}[i,j] + \frac{0.06}{|\mathcal{S}|-1}  (1-\bar{P}^{\rm bayes}[i,0]) 
\end{align}
where $\bar{P}^{\rm bayes}$ indicates the transition matrices after the Bayesian update of section \ref{subsec:updating}.
Notice that $\bar{P}$ is a probability matrix since $\sum_j \bar{P}(i,j)=1$ for each $i$.
A benefit is that this adjustment makes the submatrix $\bm Q_k$  have non-zero off-diagonal entries, making $\bar{P}$ irreducible.

\subsection{Adjustments to match propagation}

The average propagation $\rho_k$  for generation $k$  \cite{dobsonPS12} is estimated from the data   using
\begin{align}
\label{equ:rhok}
\hat\rho_{k} &= \frac{\text{Number of cascades with $> k+1 $ generations}}{\text{Number of cascades with $>k $ generations}} \notag\\
& = \frac{S(k+1)}{S(k)}  = \frac{\bm \pi_{k+1} ({\bf1} - \bm e_0)}{\bm \pi_{k} ({\bf1} - \bm e_0)}
\end{align}  
An important feature of the cascading data is that average propagation $\rho_k$ increases with generation $k$ as shown in Table~\ref{tbl:rhoemp}.
To do this, we need to form transition matrices for each of these generations that  reproduce this propagation.
\begin{table}[ht]
	\centering
	\caption{Propagations of generations $k=0$ to 17 }
	\label{tbl:rhoemp}
	\begin{tabular}{cccccccccc}
		$k$  & 0 & 1  &  2 & 3 & 4  &  5 & 6 & 7 & 8 \\
		\hline
		$\hat \rho_{k}$ & 0.13  &  0.31 & 0.44 & 0.61  & 0.73 & 0.70 & 0.78 & 0.75 & 0.71\\[2 mm]
		$k$ & 9 & 10  &  11 & 12 & 13  &  14 & 15 & 16 & 17 \\
		\hline
		$\hat \rho_{k}$ & 0.73 & 0.91 & 1.00 & 1.00  & 0.80 & 0.75 & 0.83 & 0.60 & 0.67 \\		
	\end{tabular}
\end{table}
We define a matrix $\bm A_k$ to adjust  $\bar{\bm P}_0$ and $\bar{\bm P}_{1+}$ so that the propagation  in $\bm P_{k}$ matches the empirical propagation for each generation up to generation 8. For generation 9 and above, the empirical propagation for each generation is too noisy to use individually and we combine those generations to obtain a constant transition matrix. That is, $\bm P_0 = \bar{\bm P}_{0} \bm A_0 $, $\bm P_1 = \bar{\bm P}_{1+} \bm A_1 $, ... , $\bm P_{8}= \bar{\bm P}_{1+} \bm A_{8}$, $\bm P_{9+} = \bar{\bm P}_{1+} \bm A_{9+}$.  
Then the transition matrices for all the generations are $\bm P_0, \bm P_1, \bm P_2, \bm P_3, \bm P_4, \bm P_5, \bm P_6,  \bm P_7,\bm P_8, \bm P_{9+}, \bm P_{9+}, \bm P_{9+}, ...$.

The matrix $\bm A_k$ has the effect of transferring a fraction of probability from the transient to stopping transitions and has the following form:
\begin{align}
\bm A_k=\begin{pmatrix}
1&0&...&0\\
a_k&1-a_k&...&0\\
\vdots & & \ddots &\\
a_k&0&...&1-a_k
\end{pmatrix}
\end{align} 
$a_k$ is determined from the estimated propagation rate $\hat\rho_{k}$ as follows. Using (\ref{equ:rhok}), we have
\begin{align}
\hat\rho_k&=\frac{\bm \pi_k \bar{\bm P} \bm A_k ({\bf1} - \bm e_0)}{\bm \pi_k ({\bf1} - \bm e_0)}% \notag\\
 =  (1-a_k) \frac{1-\bm \pi_k \bar{\bm P} \bm e_0}{1-\bm \pi_k \bm e_0}
 \label{akeqn}
\end{align}
and we solve (\ref{akeqn}) to obtain $a_k$ for each generation $k$.

\section{Discussion and Conclusion}

We process observed transmission line outage utility data to form a generalized influence graph and the associated Markov chain that 
statistically describe cascading outages in the data.
Successive line outages, or, more precisely, successive sets of near simultaneous line outages  in the cascading data 
correspond to transitions between nodes of the influence graph and transitions in the Markov chain.
The more frequently occurring successive line outages in the cascading data
give a stronger influence between nodes and higher transition probabilities.
The generalized influence graph introduces additional states corresponding to multiple line outages that occur nearly simultaneously.
This innovation adds a manageable number of additional states and solves some problems with previous influence graphs,
making the formation of the Markov chain clearer and more rigorous.

One of the inherent challenges of cascading  is the sparse data for large cascades.
We have used several methods to partially alleviate this when estimating the Markov chain transition matrices,
including combining data for several generations, conservatively improving estimates of stopping probabilities with an empirical Bayes method, accounting for independent outages during the cascade, and matching the observed propagation for each generation.
The combined effect of these methods is to improve estimates of the Markov chain transition matrices.
Although some individual elements of these  transition matrices are nevertheless still poorly estimated, 
what matters is the variability of the results from the Markov chain, which are the 
probabilities of small, medium and large cascades.  We assess the variability of 
these estimated probabilities with a bootstrap and find them to be estimated to a useful accuracy.
This assessment of variability is necessary for getting useful estimates of large cascade probability because large cascades are rare, and probability estimates for rare events have the potential to be so wildly variable that they are useless.

The Markov chain only models the statistics of successive transitions in the observed data. Also, there is an inherent 
limitation of not being able to account for  transitions and states not present in the observed data.
That is, the common transitions and states and some of the rarer transitions and states will be present in the data and will be represented in the Markov model, while the rarer transitions and states not present in the data will be neglected. 
However, the Markov chain can produce, in addition to the observed cascades,
combinations of the observed transitions that are different than and much more extensive than the observed cascades.
The Markov chain approximates the statistics of cascading rather than reproducing only the observed cascades.

We exploit the asymptotic properties of the Markov chain to calculate the transmission lines most involved in the propagation of larger cascades, and we show with the Markov chain that upgrading these lines can significantly reduce the probability of large cascades. 
Since a large cascade of line outages with many generations is very likely to shed substantial load, mitigating large cascades will also mitigate blackouts with large amounts of load shed.

\looseness=-1
A Markov chain driven by real data incorporates all the causes, mechanisms, and conditions of the cascading that occurred, but does not  distinguish particular causes of the interactions.
However, once the lines critical to large cascades have been identified with the influence graph, the causes related to outage of those particular lines can be identified by analyzing event logs and cause codes. 
Also, the overall impact on cascading of factors such as loading and weather can be studied by dividing the data into low and high loading or good and bad weather and forming influence graphs for each case. 

While the Markov model is driven by historical data in this paper, the Markov model is not limited to historical data. The Markov model could be driven by simulated cascades or a combination of simulated and historical cascades. Moreover, if the probabilities of specific cascading interactions between line outages are available, these probabilities could be combined into the entries of the Markov transition matrices.
The Markov chain is applied here to cascading transmission line outages, but the formulation would apply generally to process  real or simulated data for the cascading outage of components within or between networked infrastructures.

We show how to estimate the 
Markov chain from detailed outage data that is routinely collected by utilities. Being driven by observed data has some significant advantages of realism.
In particular, and in contrast with simulation
approaches, no assumptions about the detailed mechanisms of cascading 
need to made.
Since the Markov chain driven by utility data has different assumptions than simulation, 
we regard the Markov chain and simulation approaches as complementary.
The Markov chain driven by observed data offers another way to find critical lines and 
to test proposed mitigations of cascading by predicting the effect of the mitigation 
on the probabilities of small, medium, and large cascades.

\appendices

\section{Deriving the quasi-stationary distribution $\bm d_\infty$}
\label{app3}

The quasi-stationary distribution %is the stationary distribution given that the cascade is propagating and it 
can be derived in a standard way
\cite{darrochJAP65,vandoornEJOR13}.
Let $\bm d_k$ be a vector with entry $d_k[i]$ which is the probability that a cascade is in nonempty state $s_{i}$ at generation $k$ given that the cascade is propagating, that is 
\begin{align}
	d_k[i]=\frac{{\rm P}[X_k = s_i]}{{\rm P}[X_k \neq s_0]} = \frac{\pi_k[i]}{1-\pi_k[0]}, \qquad i=1,...,|\mathcal{S}|\notag
\end{align}
Then the quasi-stationary distribution is $\bm d_\infty = \lim_{k \rightarrow \infty}\bm d_{k}$.

Diagonal entries of $\bar{\bm Q}_{1+}$ corresponding to $\bar{\bm P}_{1+}$ are all zero and all other entries are positive. According to the Perron-Frobenius theorem \cite{stewartIntr94}, $\bar{\bm Q}_{1+}$ has a unique maximum modulus eigenvalue $\mu$, which is real, positive and simple with left eigenvector $\bm v'$.  By normalizing $\bm v'$, we make $\bm v'$ a probability vector. We write $\bm w$ for the corresponding right eigenvector. 
Moreover, $0<\mu<1$ and $\mu$ is strictly greater than the modulus of the other eigenvalues of $\bar{\bm Q}_{1+}$.
Suppose the cascade starts  with probability distribution $\bm \pi_0$ (note that $\pi_0[0]=0$).
According to (\ref{equ:pik}), the probability of being in state $i$ at generation $k$ is 
$\pi_k[i]= (\bm \pi_{0} \bm P_{0} \bm P_{1} ... \bm P_{k-2} \bm P_{k-1})[i] = (\bm \pi_0 \bm P^{(k-1)})[i]$.
In particular,  the probability that the cascade terminates by generation $k$ is $\pi_k[0]=\bm \pi_0 \bm P^{(k)}[0]=\bm \pi_0 \bm P^{(k)} \bm e_0$. Then for $i=1,...,|\mathcal{S}|$,
\begin{align}
 d_{k+1}[i]=\frac{\pi_{k+1}[i]}{1-\pi_{k+1}[0]}=\frac{ (\bm \pi_0  \bm P^{(k)} ) [i]}{1- \bm \pi_0 \bm P^{(k)} \bm e_0}=\frac{ (\bm \pi_0 \bm P^{(k)})[i]}{ \bm \pi_0 \bm P^{(k)}({\bf 1}- \bm e_0)}\notag
\end{align}
The first row of  $\bm P_k$ is always $[1 \quad 0 \quad ... \quad 0]$. Since $\pi_0[0]=0$, let $ \bm \pi_0 = [0 \quad \bar{ \bm \pi}_0]$. Then $\bm \pi_0 \bm P^{(k)}({\bf 1}- \bm e_0) =\bar{\bm \pi}_0 \bm Q^{(k)} {\bf 1} $ and $(\bm \pi_0 \bm P^{(k)}) [i] = (\bar{\bm \pi}_0 \bm Q^{(k)}) [i]$ for $i=1,...,|\mathcal{S}|$. And $ \bm Q^{(k)}= \bar{\bm Q}_0 \bar{ \bm Q}_{1+}^{k-1} \prod_{m=0}^{k}(1-\alpha_m)$, so that  $\bm d_{\infty}=
\lim_{k \rightarrow \infty}\bm d_{k+1}$ is
\begin{align}
\bm d_\infty&= \lim_{k\rightarrow \infty} \frac{\bar{ \bm p}_0 \bm Q^{(k)} }{\bar{ \bm p}_0 \bm Q^{(k)}{\bf 1}} = \lim_{k\rightarrow \infty}  \frac{\bar{ \bm p}_0 \bar{\bm Q}_0 \bar{ \bm Q}_{1+}^{k-1} \prod_{m=0}^{k}(1-\alpha_m)}{\bar{ \bm p}_0 \bar{\bm Q}_0 \bar{ \bm Q}_{1+}^{k-1} \prod_{m=0}^{k}(1-\alpha_m) {\bf 1}}  \notag\\
&= \frac{\bar{ \bm p}_0  \bar{\bm Q}_0 \mu^{k-1}  \bm w  \bm v'  }{\bar{ \bm p}_0  \bar{\bm Q}_0 \mu^{k-1}  \bm w  \bm v' {\bf 1} } 
 = \bm v' \notag
\end{align}
where $ \bar{\bm Q}^{(k-1)} \rightarrow  \mu^{k-1}  \bm w  \bm v' $ as $k\rightarrow\infty$.
Therefore, the dominant left eigenvector of $\bar{ \bm Q}_{1+}$ is $\bm d_\infty$.

\looseness=-1
For our data, the top three eigenvalues in modulus are $\mu=0.502$ and $-0.136\pm 0.122\,\rm i$ with corresponding moduli 0.502 and 0.381.

\bibliographystyle{IEEEtran}
\bibliography{IEEEabrv,reference1}

\begin{thebibliography}{10}
\providecommand{\url}[1]{#1}
\csname url@samestyle\endcsname
\providecommand{\newblock}{\relax}
\providecommand{\bibinfo}[2]{#2}
\providecommand{\BIBentrySTDinterwordspacing}{\spaceskip=0pt\relax}
\providecommand{\BIBentryALTinterwordstretchfactor}{4}
\providecommand{\BIBentryALTinterwordspacing}{\spaceskip=\fontdimen2\font plus
\BIBentryALTinterwordstretchfactor\fontdimen3\font minus
  \fontdimen4\font\relax}
\providecommand{\BIBforeignlanguage}[2]{{%
\expandafter\ifx\csname l@#1\endcsname\relax
\typeout{** WARNING: IEEEtran.bst: No hyphenation pattern has been}%
\typeout{** loaded for the language `#1'. Using the pattern for}%
\typeout{** the default language instead.}%
\else
\language=\csname l@#1\endcsname
\fi
#2}}
\providecommand{\BIBdecl}{\relax}
\BIBdecl

\bibitem{hinesEP09}
P.~Hines, J.~Apt, and S.~Talukdar, ``Large blackouts in {N}orth {A}merica:
  Historical trends and policy implications,'' \emph{Energy Policy}, vol.~37,
  no.~12, pp. 5249--5259, Dec. 2009.

\bibitem{carrerasPS16}
B.~A. Carreras, D.~E. Newman, and I.~Dobson, ``North {A}merican blackout time
  series statistics and implications for blackout risk,'' \emph{IEEE Trans.
  Power Syst.}, vol.~31, no.~6, pp. 4406--4414, Nov. 2016.

\bibitem{baldickPES08}
R.~Baldick, B.~Chowdhury, I.~Dobson \emph{et~al.}, ``Initial review of methods
  for cascading failure analysis in electric power transmission systems,'' in
  \emph{IEEE PES General Meeting}, Pittsburgh, PA, USA, Jul. 2008.

\bibitem{papicPES11}
M.~Papic, K.~Bell, Y.~Chen \emph{et~al.}, ``Survey of tools for risk assessment
  of cascading outages,'' in \emph{IEEE PES General Meeting}, Detroit, MI, USA,
  Jul. 2011.

\bibitem{bialekPS16}
J.~Bialek \emph{et~al.}, ``Benchmarking and validation of cascading failure
  analysis tools,'' \emph{IEEE Trans. Power Syst.}, vol.~31, no.~6, pp.
  4887--4900, Nov. 2016.

\bibitem{ciapessoniPMAPS18}
E.~Ciapessoni \emph{et~al.}, ``Benchmarking quasi-steady state cascading outage
  analysis methodologies,'' in \emph{Prob. Methods Applied to Power Syst.},
  Boise, ID, USA, Jun. 2018.

\bibitem{papicPMAPS16}
M.~Papic and I.~Dobson, ``Comparing a transmission planning study of cascading
  with historical line outage data,'' in \emph{Prob. Methods Applied to Power
  Syst.}, beijing, China, Oct. 2016.

\bibitem{AsavathirathamCSM01}
C.~{Asavathiratham}, S.~{Roy}, B.~{Lesieutre}, and G.~{Verghese}, ``The
  influence model,'' \emph{IEEE Control Syst. Mag.}, vol.~21, no.~6, pp.
  52--64, Dec. 2001.

\bibitem{Rahnamay-NaeiniCNC16}
M.~{Rahnamay-Naeini}, ``Designing cascade-resilient interdependent networks by
  optimum allocation of interdependencies,'' in \emph{Int. Conf. Computing
  Networking and Communications}, Kauai, HI, USA, Feb. 2016.

\bibitem{hinesHICSS13}
P.~Hines, I.~Dobson, E.~Cotilla-Sanchez \emph{et~al.}, ```{D}ual graph' and
  `random chemistry' methods for cascading failure analysis,'' in \emph{Proc.
  46th Hawaii Intl. Conf. System Sciences}, Maui, HI, USA, Jan. 2013.

\bibitem{qiPS15}
J.~Qi, K.~Sun, and S.~Mei, ``An interaction model for simulation and mitigation
  of cascading failures,'' \emph{IEEE Trans. Power Syst.}, vol.~30, no.~2, pp.
  804--819, Mar. 2015.

\bibitem{hinesPS17}
P.~Hines, I.~Dobson, and P.~Rezaei, ``Cascading power outages propagate locally
  in an influence graph that is not the actual grid topology,'' \emph{IEEE
  Trans. Power Syst.}, vol.~32, no.~2, pp. 958--967, Mar. 2017.

\bibitem{zhouPMAPS18}
K.~Zhou, I.~Dobson, P.~Hines, and Z.~Wang, ``Can an influence graph driven by
  outage data determine transmission line upgrades that mitigate cascading
  blackouts?'' in \emph{Prob. Methods Applied Power Syst.}, Boise, ID, USA,
  Jun. 2018.

\bibitem{qiPS18}
J.~Qi, J.~Wang, and K.~Sun, ``Efficient estimation of component interactions
  for cascading failure analysis by {EM} algorithm,'' \emph{IEEE Trans. Power
  Syst.}, vol.~33, no.~3, pp. 3153--3161, May 2018.

\bibitem{ZhangEPECS13}
X.~Zhang, F.~Liu, R.~Yao \emph{et~al.}, ``Identification of key transmission
  lines in power grid using modified k-core decomposition,'' in \emph{Proc. 3rd
  Int. Conf. Electric Power and Energy Conversion Systems}, Istanbul, Turkey,
  Oct. 2013.

\bibitem{Merrillpreprint16}
\BIBentryALTinterwordspacing
H.~M. Merrill and J.~W. Feltes, ``Cascading blackouts: Stress, vulnerability,
  and criticality,'' \emph{preprint}, 2016. [Online]. Available:
  \url{http://www.merrillenergy.com}
\BIBentrySTDinterwordspacing

\bibitem{MaSG19}
Z.~{Ma}, C.~{Shen}, F.~{Liu}, and S.~{Mei}, ``Fast screening of vulnerable
  transmission lines in power grids: A pagerank-based approach,'' \emph{IEEE
  Trans. Smart Grid}, vol.~10, no.~2, pp. 1982--1991, Mar. 2019.

\bibitem{YangPRL17}
Y.~Yang, T.~Nishikawa, and A.~E. Motter, ``Vulnerability and cosusceptibility
  determine the size of network cascades,'' \emph{Phys. Rev. Lett.}, vol. 118,
  no.~4, p. 048301, 2017.

\bibitem{CarrerasHICSS12}
B.~A. {Carreras}, D.~E. {Newman}, and I.~{Dobson}, ``Determining the
  vulnerabilities of the power transmission system,'' in \emph{Proc. 45th
  Hawaii Int. Conf. System Sciences}, Maui, HI, USA, Jan. 2012, pp. 2044--2053.

\bibitem{NakarmiNSE19}
U.~{Nakarmi}, M.~{Rahnamay-Naeini}, and H.~{Khamfroush}, ``Critical component
  analysis in cascading failures for power grids using community structures in
  interaction graphs,'' \emph{IEEE Trans. Netw. Sci. Eng.}, 2019.

\bibitem{JuJESTCS2017}
W.~{Ju}, K.~{Sun}, and J.~{Qi}, ``Multi-layer interaction graph for analysis
  and mitigation of cascading outages,'' \emph{IEEE Trans. Emerg. Sel. Topics
  Circuits Syst.}, vol.~7, no.~2, pp. 239--249, Jun. 2017.

\bibitem{ChenAccess19}
C.~Chen, W.~Ju, K.~Sun, and S.~Ma, ``Mitigation of cascading outages using a
  dynamic interaction graph-based optimal power flow model,'' \emph{IEEE
  Access}, vol.~7, pp. 168\,637--168\,648, 2019.

\bibitem{WeiPS18}
X.~Wei, J.~Zhao, T.~Huang, and E.~Bompard, ``A novel cascading faults graph
  based transmission network vulnerability assessment method,'' \emph{IEEE
  Trans. Power Syst.}, vol.~33, no.~3, pp. 2995--3000, May 2018.

\bibitem{Sunbook}
K.~Sun, Y.~Hou, W.~Sun, and J.~Qi, \emph{Power System Control Under Cascading
  Failures: Understanding, Mitigation, and System Restoration}.\hskip 1em plus
  0.5em minus 0.4em\relax Wiley-IEEE Press, 2019.

\bibitem{WangPS11}
A.~Wang, Y.~Luo, G.~Tu \emph{et~al.}, ``Vulnerability assessment scheme for
  power system transmission networks based on the fault chain theory,''
  \emph{IEEE Trans. Power Syst.}, vol.~26, no.~1, pp. 442--450, Feb. 2011.

\bibitem{LuoPESGM17}
C.~Luo, J.~Yang, Y.~Sun \emph{et~al.}, ``Identify critical branches with
  cascading failure chain statistics and hypertext-induced topic search
  algorithm,'' in \emph{IEEE PES General Meeting}, Chicago, IL, USA, 2017.

\bibitem{LiHaoWuPESGM18}
L.~{Li}, H.~{Wu}, and Y.~{Song}, ``Temporal difference learning based critical
  component identifying method with cascading failure data in power systems,''
  in \emph{IEEE PES General Meeting}, Portland, OR, USA, Aug. 2018.

\bibitem{LiHaoWuEPSR19}
L.~Li, H.~Wu, Y.~Song, and Y.~Liu, ``A state-failure-network method to identify
  critical components in power systems,'' \emph{Electric Power Systems
  Research}, vol. 181, p. 106192, 2020.

\bibitem{WangHICSS12}
Z.~{Wang}, A.~{Scaglione}, and R.~J. {Thomas}, ``A {Markov}-transition model
  for cascading failures in power grids,'' in \emph{Proc. 45th Hawaii Int.
  Conf. System Sciences}, Maui, HI, USA, Jan. 2012, pp. 2115--2124.

\bibitem{Rahnamay-NaeiniPS14}
M.~{Rahnamay-Naeini}, Z.~{Wang}, N.~{Ghani} \emph{et~al.}, ``Stochastic
  analysis of cascading-failure dynamics in power grids,'' \emph{IEEE Trans.
  Power Syst.}, vol.~29, no.~4, pp. 1767--1779, Jul. 2014.

\bibitem{Rahnamay-NaeiniSG16}
M.~{Rahnamay-Naeini} and M.~M. {Hayat}, ``Cascading failures in interdependent
  infrastructures: An interdependent {Markov}-chain approach,'' \emph{IEEE
  Trans. Smart Grid}, vol.~7, no.~4, pp. 1997--2006, Jul. 2016.

\bibitem{wangMahshidPS18}
Z.~{Wang}, M.~{Rahnamay-Naeini}, J.~M. {Abreu} \emph{et~al.}, ``Impacts of
  operators' behavior on reliability of power grids during cascading
  failures,'' \emph{IEEE Trans. Power Syst.}, vol.~33, no.~6, pp. 6013--6024,
  Nov. 2018.

\bibitem{mahshidArXiv19}
U.~Nakarmi, M.~Rahnamay-Naeini, M.~J. Hossain, and M.~A. Hasnat, ``Interaction
  graphs for reliability analysis of power grids: A survey,'' \emph{arXiv
  preprint arXiv:1911.00475 [physics.soc-ph]}, 2019.

\bibitem{dobsonPS12}
I.~Dobson, ``Estimating the propagation and extent of cascading line outages
  from utility data with a branching process,'' \emph{IEEE Trans. Power Syst.},
  vol.~27, no.~4, pp. 2146--2155, Nov. 2012.

\bibitem{bpadata}
\BIBentryALTinterwordspacing
Bonneville power administration transmission services operations \& reliability
  website. [Online]. Available:
  \url{https://transmission.bpa.gov/Business/Operations/Outages}
\BIBentrySTDinterwordspacing

\bibitem{dobsonPS16}
I.~Dobson \emph{et~al.}, ``Obtaining statistics of cascading line outages
  spreading in an electric transmission network from standard utility data,''
  \emph{IEEE Trans. Power Syst.}, vol.~31, no.~6, pp. 4831--4841, Nov. 2016.

\bibitem{davisonbootstrap97}
A.~C. Davison and D.~V. Hinkley, \emph{Bootstrap methods and their application
  (Vol. 1)}.\hskip 1em plus 0.5em minus 0.4em\relax Cambridge, U.K.: Cambridge
  Univ. Press, 1997.

\bibitem{VaimanPS12}
M.~Vaiman \emph{et~al.}, ``Risk assessment of cascading outages: methodologies
  and challenges,'' \emph{IEEE Trans. Power Syst.}, vol.~27, no.~2, pp.
  631--641, May 2012.

\bibitem{DobsonArXiv18}
I.~Dobson, ``Finding a {Z}ipf distribution and cascading propagation metric in
  utility line outage data,'' \emph{arXiv preprint arXiv:1808.08434
  [physics.soc-ph]}, 2018.

\bibitem{Guikema2007Formulating}
S.~D. Guikema, ``Formulating informative, data-based priors for failure
  probability estimation in reliability analysis,'' \emph{Reliability
  Engineering \& System Safety}, vol.~92, no.~4, pp. 490--502, Apr. 2007.

\bibitem{carlinBayesian08}
B.~P. Carlin and T.~A. Louis, \emph{Bayesian methods for data analysis}.\hskip
  1em plus 0.5em minus 0.4em\relax Boca Raton, FL, USA: CRC Press, 2008.

\bibitem{jaynes86}
E.~T. Jaynes, ``Bayesian methods: General background,'' in \emph{Maximum
  Entropy and Bayesian Methods in Applied Statistics, Cambridge, U.K.}\hskip
  1em plus 0.5em minus 0.4em\relax Cambridge Univ. Press, 1986.

\bibitem{fangEntropy12}
S.-C. Fang, J.~R. Rajasekera, and H.-S.~J. Tsao, \emph{Entropy optimization and
  mathematical programming}.\hskip 1em plus 0.5em minus 0.4em\relax Springer,
  2012.

\bibitem{darrochJAP65}
J.~N. Darroch and E.~Seneta, ``On quasi-stationary distributions in absorbing
  discrete-time finite {M}arkov chains,'' \emph{J. Appl. Probab.}, vol.~2,
  no.~1, pp. 88--100, Jun. 1965.

\bibitem{vandoornEJOR13}
E.~V. Doorn and P.~Pollett, ``Quasi-stationary distributions for discrete-state
  models,'' \emph{European J. Operat. Res.}, vol. 230, pp. 1--14, 2013.

\bibitem{stewartIntr94}
W.~J. Stewart, \emph{Introduction to the numerical solution of Markov
  chains}.\hskip 1em plus 0.5em minus 0.4em\relax Princeton, NJ, USA: Princeton
  Univ. Press, 1994.

\end{thebibliography}

{\bf Kai Zhou} (S'19) received the B.S degree in electrical engineering from China Agriculture University in 2014, and the M.S. degree in electrical engineering from Tianjin University in 2017. He is currently pursuing the Ph.D. degree with Iowa State University, Ames, IA, USA. His current research interests include cascading failures and data analytics.

{\bf Ian Dobson} (F'06) received the B.A. degree in mathematics from Cambridge University, UK, and the Ph.D. degree in electrical engineering from Cornell University, USA.
He is currently Sandbulte Professor of Engineering at Iowa State University, Ames, IA, USA.

{\bf Zhaoyu Wang} (S'13-M'15) is the Harpole-Pentair Assistant Professor with Iowa State University. He received the B.S. and M.S. degrees in electrical engineering from Shanghai Jiaotong University in 2009 and 2012, respectively, and the M.S. and Ph.D. degrees in electrical and computer engineering from Georgia Institute of Technology in 2012 and 2015, respectively. His research interests include power distribution systems and microgrids, particularly on their data analytics and optimization. He is the Principal Investigator for a multitude of projects focused on these topics and funded by the National Science Foundation, the Department of Energy, National Laboratories, PSERC, and Iowa Energy Center. Dr. Wang is the Secretary of IEEE Power and Energy Society (PES) Award Subcommittee, Co-Vice Chair of PES Distribution System Operation and Planning Subcommittee, and Vice Chair of PES Task Force on Advances in Natural Disaster Mitigation Methods. He is an editor of IEEE Transactions on Power Systems, IEEE Transactions on Smart Grid, IEEE PES Letters and IEEE Open Access Journal of Power and Energy, and an associate editor of IET Smart Grid.

{\bf Alexander Roitershtein} is a Research Associate at the Department of Statistics, Texas A\&M University. He received the Ph.D. in Applied Mathematics from Technion, Israel Institute of Technology in 2004. His research interests include random walk models and their applications, population dynamics, complex networks, microbiome ecology, applications of branching processes to biology, stochastic processes in random environment, and general theory of Markov chains.

{\bf Arka Ghosh} is a Professor of Statistics at Iowa State University and holds courtesy appointments in Departments of Mathematics as well as in Industrial Manufacturing and Systems Engineering. He received his B. Stat and M. Stat degrees from Indian Statistical Institute and Ph.D in Statistics from University of North Carolina at Chapel Hill and joined Iowa State University in 2005. His interests are primarily in Applied Probability and Statistics, including Stochastic Modeling, Queueing theory, Random Graphs, Network Modeling.

\end{document}